# National and International Dimensions of the Triple Helix in Japan:

## University-Industry-Government *versus* International Co-Authorship Relations

Loet Leydesdorff[*] and Yuan Sun[**]

*Amsterdam School of Communications Research (ASCoR),
University of Amsterdam, Kloveniersburgwal 48, 1012 CX Amsterdam, The Netherlands;
loet@leydesdorff.net; http://www.leydesdorff.net

**National Institute of Informatics (NII), 2-1-2 Hitotsubashi, Chiyoda-ku,
Tokyo 101-8430, Japan; yuan@nii.ac.jp

**Abstract**

International co-authorship relations and university-industry-government ("Triple Helix") relations have hitherto been studied separately. Using Japanese (ISI) publication data for the period 1981-2004, we were able to study both kinds of relations in a single design. In the Japanese file, 1,277,823 *articles* with at least one Japanese address were attributed to the three sectors, and we know additionally whether these papers were co-authored internationally. Using the mutual information in three and four dimensions, respectively, we show that the Japanese Triple-Helix system has continuously been eroded at the national level. However, since the middle of the 1990s, international co-authorship relations have contributed to a reduction of the uncertainty. In other words, the national publication system of Japan has developed a capacity to retain surplus value generated internationally. In a final section, we compare these results with an analysis based on similar data for Canada. A relative uncoupling of local university-industry relations because of international collaborations is indicated in both national systems.

**Introduction**

The formation of research teams leading to co-authored publications has been perhaps the major development in the sciences in recent decades (DeBeaver & Rosen, 1978; Luukkonen *et al.*, 1992, 1993; Gibbons *et al.*, 1994; Hicks & Katz, 1996). International co-authorship relations have grown exponentially during the past two decades (Zitt *et al.*, 2000; Glänzel, 2001; Wagner & Leydesdorff, 2005; Wagner, 2008). Co-authored publications across sectors of the economy (university, industry, government) have been used as indicators of the Triple Helix model (Leydesdorff, 2003), and more generally for the study of science-technology relations (Narin *et al.*, 1997) and entrepreneurial science (Tijssen, 2006).

Indicators of university-industry relations, on the one hand, have mainly focused on the national level because of their potential relevance for government policies (Lundvall, 1988; Nelson *et al.*, 1993). Danell & Persson (2003) further decomposed the Swedish national systems of publications, patents, and persons ("Triple P") in terms of regions using the information provided by postal codes. On the other hand, the study of international co-authorship relations has been pursued mainly at the global level, or sometimes also disaggregated in terms of disciplines (e.g., Wagner & Leydesdorff, 2003). The combination of the two types of studies, that is, studies of international collaboration and triple-helix collaborations, was hitherto not possible because this requires a huge amount of data processing and cleaning of the institutional address information. Within each country, even the attribution of addresses to institutes is already a non-trivial task, let alone the identification of all institutes in terms of sectors in the economy.

This report builds on the ongoing work of a Japanese team (Sun *et al.*, 2006) to categorize all publications with at least one Japanese address included during the period 1981-2004 in the *Science Citation Index,* the *Social Science Index*, and the *Arts & Humanities Index* in terms of their precise origins. In addition to information about domestic university-industry-government relations, this database also indicates international co-authorship relations. Thus, the data can be used for the measurement of the Triple Helix model in terms of the mutual information among three or four dimensions. Additionally, the interaction between the national and international levels can be specified and appreciated.

The mutual information among three or more dimensions provides us with (Shannon-type) information about the uncertainty that prevails in a network. For example, it matters for a government whether or not universities and industries already entertain strong relations. If the indicator is positive, the uncertainty is increased, while a negative value indicates reduction of the uncertainty. One can consider this negative uncertainty also as an indicator of synergy. The synergy in the configuration cannot be attributed to any of the contributing partners: the intensity of their relations and the size of their contributions matter for this structural effect. The research question is whether such a reduction of uncertainty is indicated in the relations among the different dimensions, and if so, which dimensions are more important for this effect than others (Leydesdorff and Fritsch, 2006; Leydesdorff *et al.*, 2006; Leydesdorff, 2006)?

Because the results of our analysis (to be presented below) were surprising, we asked research teams in several countries whether similar attribution efforts of institutional addresses had been made in their respective national centers. While a number of research teams claimed to have this cleaned data at hand, we only obtained the Canadian data collected by the *Observatoire des* *sciences et technologies* in Montreal (Godin & Gingras, 2000). However, the Canadian data is based on the CD-Rom version of the database, while the Japanese data were commissioned at Thomson/ISI as a so-called *National Citation Report* (for Japan). The Canadian file contains articles, reviews, and notes included in the *Science Citation Index* (Larivière *et al*., 2006), while the Japanese data contains only articles in the three citation databases combined, that is, including the *Social Science Citation Index* and the *Arts & Humanities Index*. Both databases cover the same period (1981-2004).

Because of these differences in the data, one should not consider this as a comparative study. We used the Canadian data as a background check. The comparison allows us to focus on main trends and to test whether these are robust despite differences in the data types.

**Methods**

The mutual information in two dimensions (or transmission *T*) follows directly from the Shannon formulas. *T* is defined as the difference which it makes when two probability distributions are combined:

$$H_i = - \Sigma_i \, p_i \, \log_2 (p_i); \quad H_{ij} = - \Sigma_i \, \Sigma_j \, p_{ij} \, \log_2 (p_{ij})$$

$$H_{ij} = H_i + H_j - T_{ij}$$

$$T_{ij} = H_i + H_j - H_{ij} \quad (1)$$

$T_{ij}$ is zero if the two distributions are completely independent, but otherwise necessarily positive (Theil, 1972). Abramson (1963, at p. 129) derived that the mutual information in three

dimensions—let us use "u" for university, "i" for industry, and "g" for government—can be defined analogously as follows:

$$T_{uig} = H_u + H_i + H_g - H_{ui} - H_{ug} - H_{ig} + H_{uig} \qquad (2)$$

The resulting indicator can be negative or positive (or zero) depending on the relative sizes of the contributing terms. A negative value means that the uncertainty prevailing at the network level is reduced. McGill (1954) proposed calling this negative uncertainty "configurational information" (Jakulin & Bratko, 2004). Because the information is configurational, the reduction of the uncertainty cannot be attributed to one of the contributors; these network effects are systemic.

The extension with a fourth dimension ("f" for foreign) is straightforward:

$$T_{uigf} = H_u + H_i + H_g + H_f - H_{ui} - H_{ug} - H_{uf} - H_{ig} - H_{if} - H_{gf}$$
$$+ H_{uig} + H_{uif} + H_{ugf} + H_{igf} - H_{uigf} \qquad (3)$$

All distributions needed for the computation are contained in the data, except for the number of foreign papers *without* a Japanese address. However, this number is needed for the computation of $H_f$. Instead of assuming that all non-Japanese publications in the international database are relevant for the Japanese system, we add the total of all papers in the set with at least one non-Japanese address as a proxy for the *relevant* non-Japanese environment. For reasons of proper normalization, the sum total of papers is increased with this number in the international case study.

**Data**

During the period 1981-2004 a total of 1,453,888 papers with at least one Japanese address were published in the three above-mentioned ISI databases. The study is based on the subset of 1,277,823 articles among this data (87.9%). The records were standardized in terms of precise addresses, and where necessary, name changes because of mergers and acquisitions were corrected so that a set of unique address identifiers could be generated. In many cases, this implied also the unification of different English translations of the same Japanese name (Sun *et al.*, forthcoming).

As in most OECD countries, university addresses are involved in a high percentage of these papers. During the 1980s, this percentage decreased by one percent point from 80.5% to 79.5%, but thereafter it increased continuously to 85.2% in 2004. However, the percentage of articles with only domestic university addresses declined steadily from 69.4% in 1991 to 45.3% in 2004 as university scholars became much more involved in collaborations.

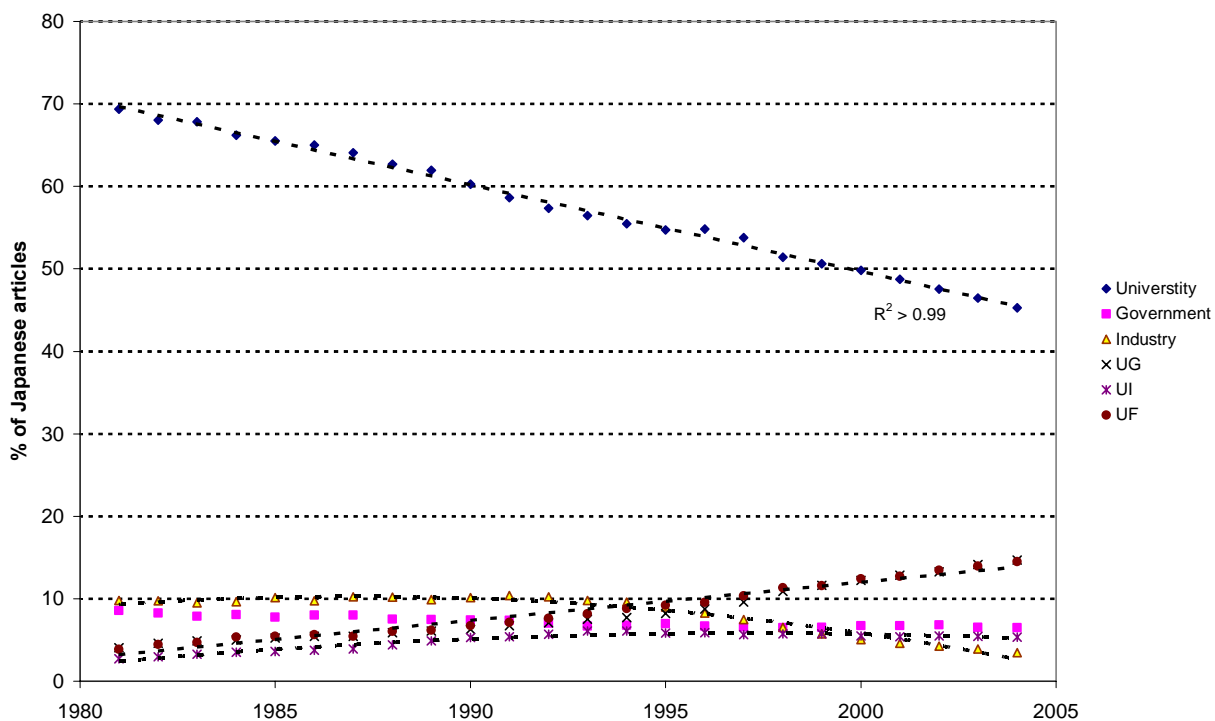

**Figure 1**: Co-authorship relations in the Japanese set during the period 1981-2004; series with values lower than 3% are not shown (IG, IF, GF, UIF, UGF, IGF, UIG, UIGF).

The shares of both internationally co-authored papers and papers co-authored between university research and research in public-sector laboratories increased. The percentages are almost identical: 4.0 (± 0.1) % in 1981 to 14.6 (± 0.1) % in 2004. However, the percentage of articles with exclusively an industrial address decreased dramatically during the 1990s: from 10.4% in 1991 to 3.5% in 2004. This is not compensated by university-industry relations because this percentage also decreased during the 1990s albeit by less than one percentage point.

The lines penciled into Figure 1 are drawn in order to assist the reader with the interpretation. They are based on linear regression in the case of straight lines and on second-order polynomials in the case of bending curves. (All fits are larger than .95.) The other series are all below three percent of the total publication volume. For example, the percentage papers with addresses in all three sectors (UIG) has increased from 0.2% in 1981 to 1.6% in 2004. The number of papers with all three sectors involved and international co-authorship was 168 (0.3%) in 2004 as against only two such co-authorships in 1981.

In summary, these descriptive statistics indicate decreasing triple-helix relationships in the Japanese publication system since 1990 because of a decline in university-industry relations in terms of co-authored publications. Until 1995, industry was the second largest producer of academic papers in Japan. The number of papers incorporating both industrial and government addresses has remained low both with and without university participation.

**Results**

*a. The Japanese publication system*

Let us first turn to the mutual information in three dimensions for the Japanese national system of publications, operationalized as a Triple Helix of university-industry-government relations. In the national case, the solution is exact, since the assumption about the number of foreign papers is not yet needed.

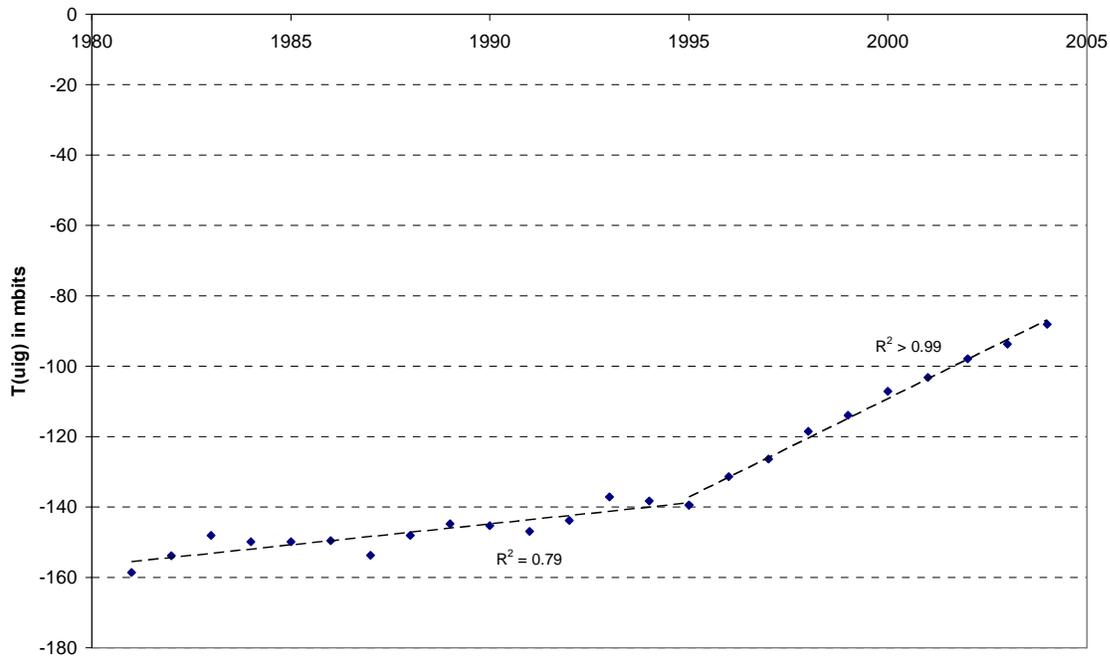

**Figure 2**: The mutual information in three dimensions of Japanese articles with addresses in university, industry, and government establishments, and their co-authorship relations.

Figure 2 shows a trend breach around 1995. Although a decline in the participation of industry in the national portfolio was noted above upon visual inspection of Figure 1, the loss of synergy within the system is caused not only by this decline, but also by the erosion of university-industry co-authorship relations within the system (Figure 3).

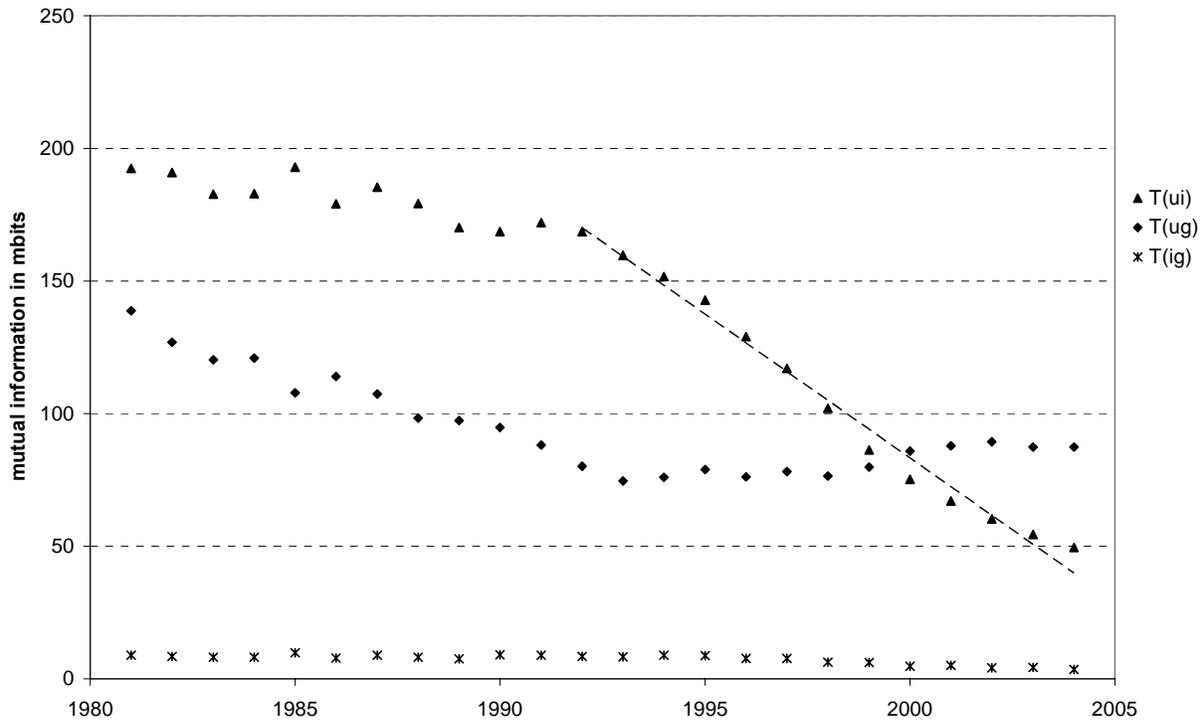

**Figure 3**: Mutual relations between university, industry, and government addresses in the domain of articles with Japanese addresses.

The uncoupling between the academic and industrial publication systems began already before 1995, but is reflected in the national system (Figure 2) with some delay. Note that the mutual information between universities and government agencies is approximately stable during the same period (1993-2004). The mutual information between industry and government is of a lower order of magnitude during the entire period under study.

*b. Inclusion of international co-authorship relations*

The addition of the international dimension changes all distributions because the sum total (*N*) is changed when the international environment is included. As noted, we used the sum of Japanese papers with an international co-author as a proxy for the relevant international environment.

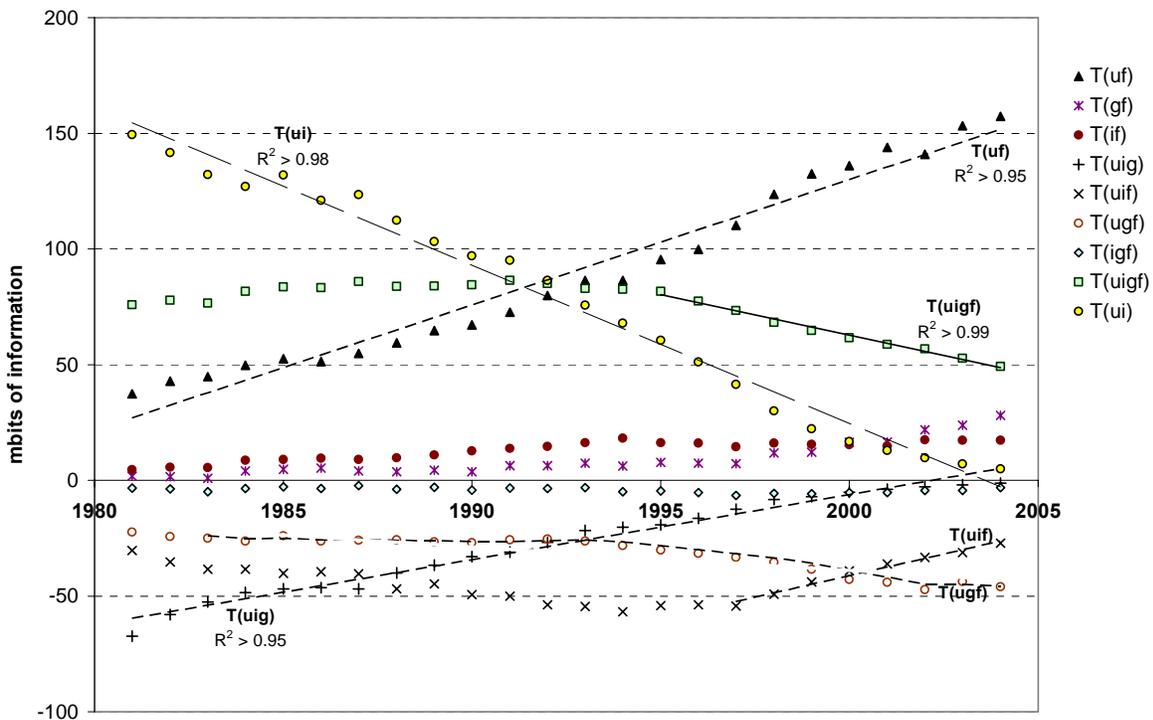

**Figure 4**: The mutual information in two, three, and four dimensions among Japanese articles with a university, industrial, or governmental address, and international co-authorships.

The upper half of Figure 4 shows two major trends in the bilateral relations. (These two trends are negatively correlated; $r = -0.99$; $p < 0.01$.) First, there is the ongoing expansion of collaboration between university researchers in Japan with international colleagues. This upward trend line (for $T_{uf}$) is indicated in the figure with ▲. The international co-authorship relations are at the systems level to the detriment of university-industry relations (indicated in this figure with ○ and as a negative trend). Both trends are robust and take place over the full period of time (1981-2004).

These long-term—i.e., cultural—trends have a negative influence on the integration of the national system. This is exhibited in the lower part of the figure: the mutual information in university-industry-government relations ($T_{uig}$; indicated with a +) becomes continuously less negative during the entire period. Both $T_{uig}$ and $T_{ui}$ become approximately zero at the turn of the millennium. Thus, uncertainty in the system increases because of the erosion of both university-industry co-authorship relations and industry participation in the system in general.

The other mutual informations in three dimensions are pictured in the lower part of the graph; they are all negative, that is, reducing the uncertainty which prevails.[1] Among these trilateral systems, university-government co-authorship relations with international participation have notably gained momentum ($T_{ugf}$). Since 1995, uncertainty in the total system of university-industry-government-foreign co-authorship relations ($T_{uigf}$; indicated with □ and a solid line) has been steadily reduced. In other words, the international co-authorship relations of the national partners have increasingly been integrated into the national system, and this has reduced the uncertainty. The synergy in the configuration can no longer be found at the national level; one has to take the international dimension of the publication system into account.

---

[1] As noted, the values in Figures 2 and 3 cannot be compared directly because the normalization for the total (*N*) is different; a proxy for the total of foreign publications was added in the international dimension.

*c. Comparison with Canadian data*

As specified above, the Canadian data are different from the Japanese data in some respects, but in many ways they are comparable. Therefore, we did not focus on minor differences, but on trends which seem robust in both sets. For example, university addresses are dominant in both the Canadian and Japanese sets. In the Canadian case, the percentage of papers with university addresses increased from 82.4% in 1980 to 90.7% in 2004. The growth in international co-authorship relations among academics is much stronger in Canada than in Japan.

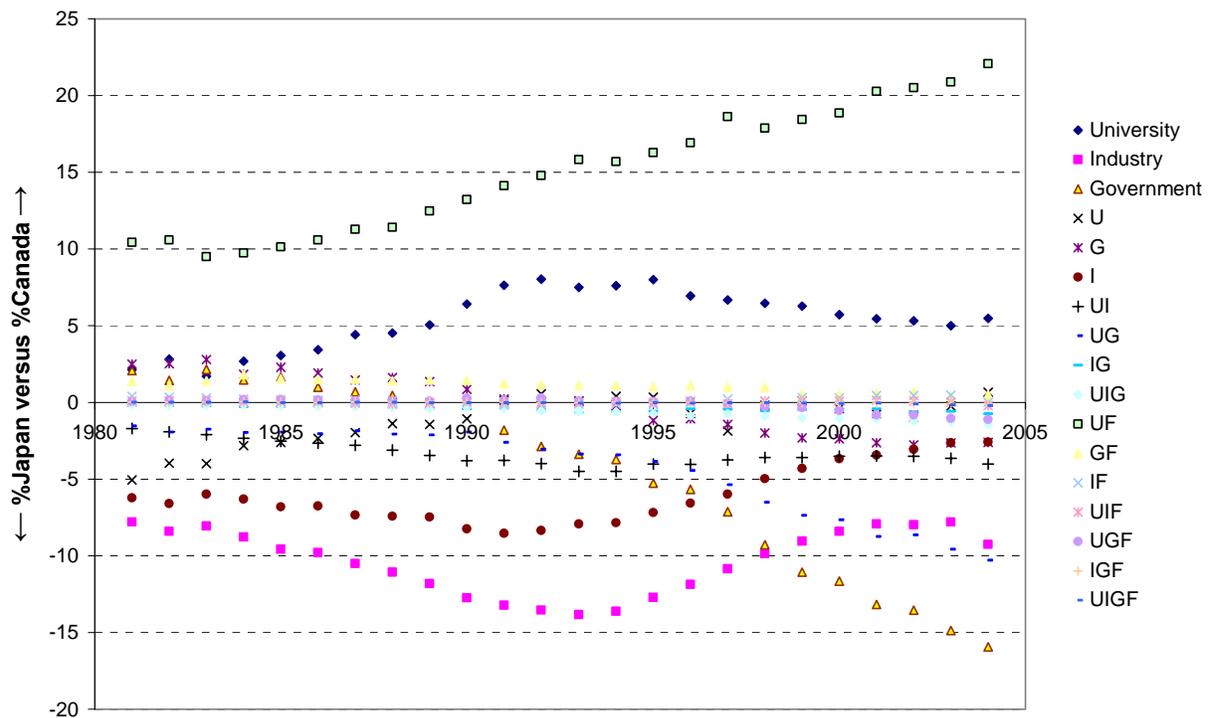

**Figure 5**: Differences in national co-authorship portfolios between Canada and Japan.

In Figure 5, the percentages of contributions to the national totals are plotted for the various categories in terms of the differences between Canada and Japan. For example, the series at the top of the figure indicates that the Canadian percentage of internationally co-authored papers with university addresses was 10% larger in the early 1980s, and this difference increased to a 22.1% in 2004. Actually, the level of these international co-authorship relations reached the level of 14.6% in Japan in 2004, while Canada had 14.3% of these papers in 1981. (The turning point in the Japanese system was reached at a level of 7-8% for this indicator during the period 1992-1995.)

The second major difference can be found in the bottom half of the picture: the percentage of publications with only governmental addresses or only industrial addresses in the Japanese set is ten percent greater than in the Canadian set. As was noted above, industrial publications in Japan declined during the 1990s, while the government share increased. Despite the decline in university-industry relations in Japan, the percentage of papers co-authored between industry and academia is higher than in Canada during the entire period. In 2004, this percentage is only 1.3% for Canada as against 5.4% for Japan.

In summary, the Canadian publication system is far more internationalized than the Japanese, but national Triple Helix relations are much stronger in the Japanese system. This result accords with received wisdom about these two systems: the Canadian publication system is highly integrated with the Anglo-Saxon one, and the labor market for academics is not constrained by the US/Canadian border. The Japanese system has lagged in terms of internationalization (Wagner, 2008), but has been strong in terms of its national integration (Irvine & Martin, 1984; Leydesdorff, 2003).

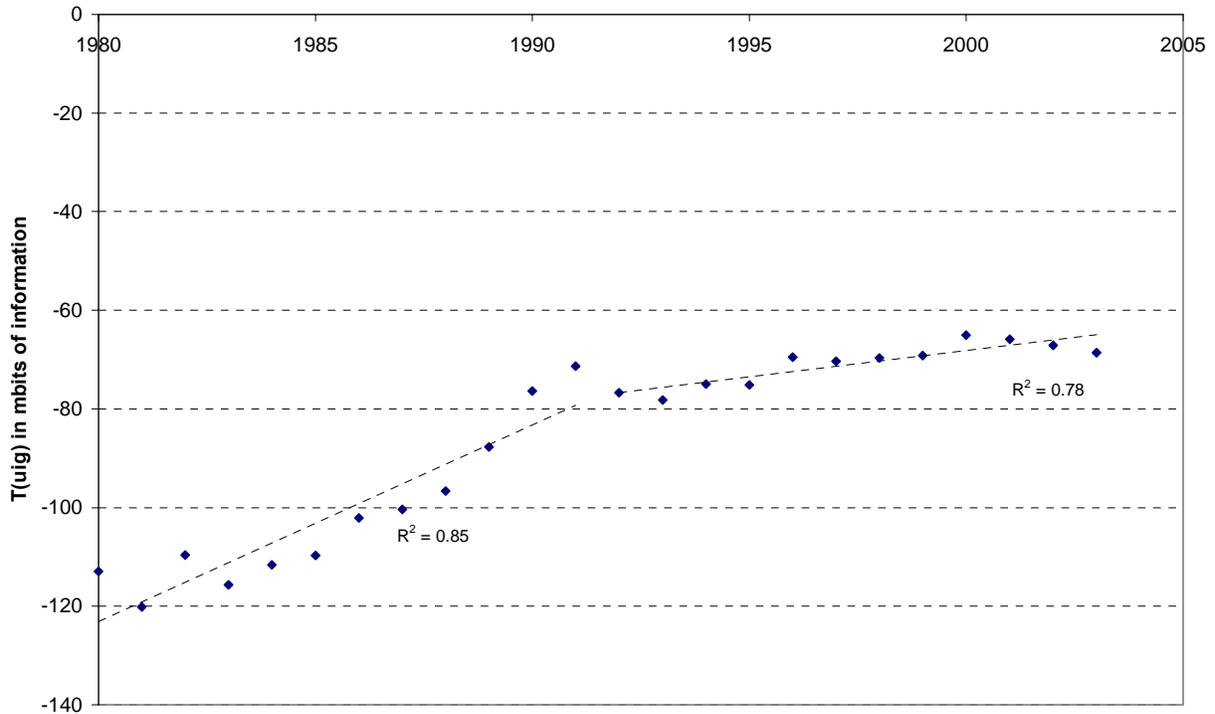

**Figure 6**: Mutual information in three dimensions of Canadian papers with addresses in university, industry, and government establishments, and their co-authorship relations.

Figure 6 shows the equivalent of Figure 2, but this time for the Canadian data. There is a similar trend breach earlier in the 1990s, and in the opposite direction. More detailed analysis reveals that the mutual information in university-industry relations no longer declined after 1990. The continuing erosion of Triple Helix relations is due to the steady decrease of university-government relations during the entire period under study. In 2004, the reduction of uncertainty in the system by university-government relations is approximately 50% of that in 1981.

The crucial comparison for this research is contained in Figure 7, which provides the analogon of Figure 4 above, but this time for the Canadian set:

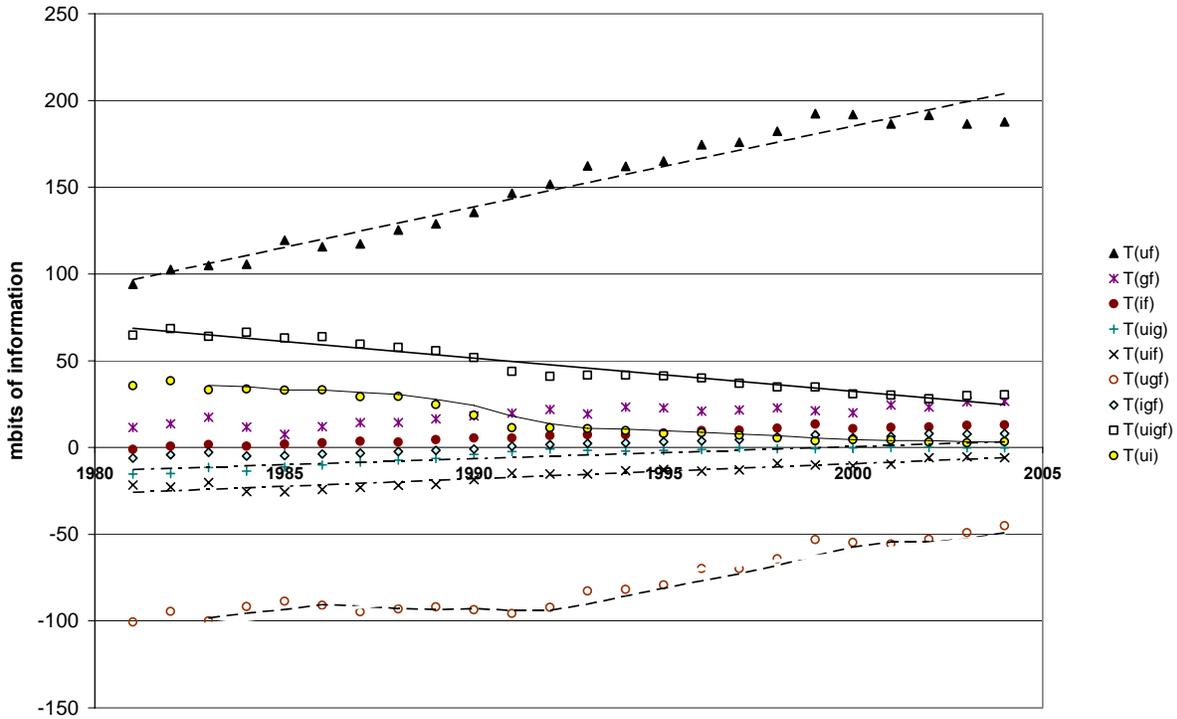

**Figure 7**: The mutual information in two, three, and four dimensions among Canadian papers with a university, industrial, or governmental address, and international co-authorships.

In addition to the continuous increase in the mutual information in international academic co-authorship relations ($T_{uf}$; indicated as ▲), this figure shows that the $T_{uigf}$ (indicated as □ and with a solid line) decreases during the entire period under study. The synergy noted above as a new phenomenon in the Japanese set in the 1990s was steady in place in the Canadian system during this whole period. The Canadian system has continuously improved its capacity to retain reduction of the uncertainty from a synergy of national and international co-authorship relations.

**Conclusions**

Using the operationalization in terms of transmissions between and among the helices, the Triple Helix metaphor can be elaborated into an indicator of the dynamics at work within a knowledge-based system (Etzkowitz & Leydesdorff, 2000; Leydesdorff & Fritsch, 2006). The Shannon-type entropies, however, are based on probability distributions whose dimensionality can be extended further. For example, the international dimension can be endogenized into the Triple Helix model as a fourth dimension.

The results suggest that long-term trends of internationalization among Japanese academic scholars in the period 1992-1995 led to the opening of the national system to the international dimension. As an OECD member, Japan can be considered as atypical because it was weakly internationalized in the early 1990s (Wagner, 2008). In the Canadian data, however, we found the international dimension as a source of synergy during the entire period under study (1980-2004).

The relevant context is provided by the globalization of the knowledge-based economy during the first half of the 1990s. This development was induced by the demise of the Soviet Union, the unification of Germany, and the opening of China after 1991 (Zhou & Leydesdorff, 2006). For this reason, one can expect similar synergies as in Japan to have emerged in other countries in Asia and in the accession countries of Eastern Europe since 1990 (Lengyel & Leydesdorff, 2007). Using these information measures, one is able to indicate such a transition of a (publication) system in terms of its relevant subdynamics.

**Acknowledgments**

The authors are grateful to Yves Gingras, scientific director of the *Observatoire des sciences et des technologies* (OST) for the Canadian data, and to Masamitsu Negishi for his help in organizing the Japanese data.

**References**

Abramson, N. (1963). *Information Theory and Coding*. New York, etc.: McGraw-Hill.

DeBeaver, D., & Rosen, R. (1978). Studies in Scientific Collaboration. Part III. Professionalization and the Natural History of Modern Scientific Co-Authorship. *Scientometrics,* 1(3), 231-245.

Danell, R., & Persson, O. (2003). Regional R&D activities and interaction in the Swedish Triple Helix. *Scientometrics,* 58(2), 205-218.

Etzkowitz, H., & Leydesdorff, L. (2000). The Dynamics of Innovation: From National Systems and 'Mode 2' to a Triple Helix of University-Industry-Government Relations. *Research Policy,* 29(2), 109-123.

Gibbons, M., Limoges, C., Nowotny, H., Schwartzman, S., Scott, P., & Trow, M. (1994). *The new production of knowledge: the dynamics of science and research in contemporary societies*. London: Sage.

Glänzel, W. (2001). National characteristics in international scientific co-authorship relations. *Scientometrics,* 51, 69-115.

Godin, B., & Gingras, Y. (2000). The place of universities in the system of knowledge production. *Research Policy,* 29(2), 273-278.

Hicks, D., & Katz, J. S. (1996). Science policy for a highly collaborative science system. *Science and Public Policy,* 23, 39-44.

Irvine, J., & Martin, B. R. (1984). *Foresight in Science: Picking the Winners*. London: Frances Pinter.

Jakulin, A., & Bratko, I. (2004). Quantifying and Visualizing Attribute Interactions: An Approach Based on Entropy. from http://arxiv.org/abs/cs.AI/0308002

Larivière, V., Gingras, Y., & Archambault, É. (2006). Canadian collaboration networks: A comparative analysis of the natural sciences, social sciences and the humanities. *Scientometrics,* 68(3), 519-533.

Lengyel, B., & Leydesdorff, L. (2007). *Measuring the knowledge base in Hungary: Triple Helix dynamics in a transition economy.* Paper presented at the 6th Triple Helix Conference, 16-19 May 2007, Singapore.

Leydesdorff, L. (2003). The Mutual Information of University-Industry-Government Relations: An Indicator of the Triple Helix Dynamics. *Scientometrics,* 58(2), 445-467.

Leydesdorff, L. (2006). *The Knowledge-Based Economy: Modeled, Measured, Simulated*. Boca Raton, FL: Universal Publishers.

Leydesdorff, L., & Fritsch, M. (2006). Measuring the Knowledge Base of Regional Innovation Systems in Germany in terms of a Triple Helix Dynamics. *Research Policy,* 35(10), 1538-1553.

Leydesdorff, L., Dolfsma, W., & Panne, G. v. d. (2006). Measuring the Knowledge Base of an Economy in terms of Triple-Helix Relations among 'Technology, Organization, and Territory'. *Research Policy,* 35(2), 181-199.

Lundvall, B.-Å. (1988). Innovation as an interactive process: from user-producer interaction to the national system of innovation. In G. Dosi, C. Freeman, R. Nelson, G. Silverberg & L. Soete (Eds.), *Technical Change and Economic Theory* (pp. 349-369). London: Pinter.

Luukkonen, T., Persson, O., & Sivertsen, G. (1992). Understanding Patterns of International Scientific Collaboration. *Science, Technology and Human Values* 17, 101-126.

Luukkonen, T., Tijssen, R. J. W., Persson, O., & Sivertsen, G. (1993). The Measurement of International Scientific Collaboration. *Scientometrics,* 28(1), 15-36.

McGill, W. J. (1954). Multivariate information transmission. *Psychometrika,* 19(2), 97-116.

Narin, F., Hamilton, K. S., & Olivastro, D. (1997). The increasing link between U.S. technology and public science. *Research Policy,* 26(3), 317-330.

Nelson, R. R. (Ed.). (1993). *National Innovation Systems: A comparative analysis*. New York: Oxford University Press.

Sun, Y., Negishi, M., Nishizawa, M., & Watanabe, K. (2006). Research Linkage between University and Industry in Japan: Comparison based on NCR-J and CJP databases. *Journal of Japan Society of Information and Knowledge,* 16(2), 7-12.

Sun, Y., Negishi, M., & Nisizawa, M. (forthcoming). Co-authorship Linkages between Universities and Industry in Japan. *Research Evaluation* (in print).

Theil, H. (1972). *Statistical Decomposition Analysis*. Amsterdam/ London: North-Holland.

Tijssen, R. J. W. (2006). University-industry interactions and university entrepreneurial science: towards measurement models and indicators. *Research Policy,* 35(10), forthcoming.

Wagner, C. S. (2008). *The New Invisible College*. Washington, DC: The Brookings Press.

Wagner, C. S., & Leydesdorff, L. (2003). Seismology as a dynamic, distributed area of scientific research. *Scientometrics,* 58(1), 91-114.

Wagner, C. S., & Leydesdorff, L. (2005). Mapping the Network of Global Science: Comparing International Co-authorships from 1990 to 2000. *International Journal of Technology and Globalization,* 1(2), 185-208.

Zhou, P., & Leydesdorff, L. (2006). The emergence of China as a leading nation in science. *Research Policy,* 35(1), 83-104.

Zitt, M., Bassecoulard, E., & Okubo, Y. (2000). Shadows of the past in international cooperation: Collaboration profiles of the top five producers of science. *Scientometrics,* 47(3), 627-657.